# Thermal Performance of a Liquid-cooling Assisted Thin Wickless Vapor Chamber

Arani Mukhopadhyay, Anish Pal, Mohamad Jafari Gukeh, Constantine M. Megaridis*
*Mechanical and Industrial Engineering, University of Illinois Chicago, IL, United States*
\* cmm@uic.edu

*Abstract*—The ever-increasing need for power consumption in electronic devices, coupled with the requirement for thinner size, calls for the development of efficient heat spreading components. Vapor chambers (VCs), because of their ability to effectively spread heat over a large area by two-phase heat transfer, seem ideal for such applications. However, creating thin and efficient vapor chambers that work over a wide range of power inputs is a persisting challenge. VCs that use wicks for circulating the phase changing media, suffer from capillary restrictions, dry-out, clogging, increase in size and weight, and can often be costly. Recent developments in wick-free wettability patterned vapor chambers, that replace traditional wicks with laser-fabricated wickless components, hold promise. An experimental setup is developed which allows for fast testing and experimental evaluation of water-charged VCs with liquid-assisted cooling. The sealed chamber can maintain vacuum for long durations, and can be used for testing of very thin wick-free VCs. This work extends our previous study by decreasing the overall thickness of the wick-free VC down to 3 mm and evaluates its performance characteristics. Furthermore, the impact of wettability patterns on VC performance is investigated, by carrying out experiments both in non-patterned and patterned VCs. Experiments are first carried out on a wick-free VC with no wettability patterns and comprising of an entirely superhydrophilic evaporator coupled with a hydrophobic condenser. Thereafter, wettability patterns that aid the rapid return of water to the heated site on the evaporator and improve condensation on the condenser of the vapor chamber are implemented. The thermal characteristics of the VCs are compared, and show that the patterned VCs outperform the non-patterned VCs under all scenarios. The patterned VCs exhibit low thermal resistance independent of fluid charging ratio, while withstanding higher power inputs without thermal dry-outs.

*Keywords—Wick free, vapor chamber, wettability engineering, wettability patterns, heat spreading*

## I. Introduction

The ever-increasing need for smaller, faster, and more powerful electronic components, has created a problem of thermal management involving removal of high heat fluxes [1]. Traditional methods of cooling viz. liquid cooling or forced convection cooling, are not sufficient to remove heat from small, constricted spaces, and hinders the development of high-power electronic packages, especially with power densities greater than $10^7$ W/m$^2$ [2].

Passive heat transfer devices, like heat pipes or vapor chambers, that work by leveraging capillary forces, can be used for rapidly transferring or spreading the heat, to reduce local heat flux, and thereby, enhance any conventional cooling process. Vapor chambers (VCs) are passive heat spreading devices, that contain a hermetically sealed vapor space, within which a working fluid is contained, generally at a lower pressure. On being exposed to heat, the working fluid inside the vapor chamber evaporates by taking in the latent heat, flows through a short distance, and collects again on the condenser side of the VC. Such rapid evaporation and condensation cycles inside the VC, which generally take place in the absence of non-condensable gases (NCGs), help rapidly transport heat from a localized section over a larger condenser, thus lowering the local heat flux. VCs with the ability to spread heat over more than one dimensions, are generally preferred over conventional heat pipes, especially in applications requiring higher heat fluxes (> 50 W/cm$^2$) [3].

Extensive research has been dedicated into development, characterization, and optimization of VC designs for specialized applications. A vapor chamber generally employs a wick for transport of the working fluid from the condenser to the evaporator; the wick also functions to form a uniform thin layer of the working fluid over the heated area, thereby minimizing thermal resistance, and raising evaporation rates. However, flow of the working fluid in a VC is generally limited by the capillary restrictions of the wick (governed by Darcy's law [1]), and can therefore lead to wick pore clogging or thermal dry-outs. VC performance and thermal resistance can be improved by further optimization of the VC wick structure or via implementation of grooved structures, which can improve heat transfer by spreading out heat in all directions away from the heater [4]. On the other hand, VC designs have been implemented with no wicks inside the chambers. Wick-free vapor chambers, or hybrid systems which include both wick free and wicked components, try to bypass such capillary restrictions by implementing wettability patterns, which can be used for pumpless transport of the working fluid [5]. Extensive studies on such wettability-patterned surfaces have been carried out earlier and they have been shown to enhance condensation heat transfer rates [6]. Wettability patterns often incorporate diverging superhydrophilic tracks on a hydrophobic or superhydrophobic surrounding surface. Droplets confined on such tracks move to the widening side of the tracks by Laplace-pressure driven spreading [7]. Such patterns have also been incorporated in design of hybrid vapor chambers featuring wickless condensers and wicked evaporators [8]. Applications of such hybrid vapor chambers even extend to air cooling of high frequency power electronics components [9, 10]. Damoulakis and Megaridis [11] demonstrated the first vapor chamber comprising entirely of wick-free components. The developed VC featured complementary wettability patterns on both the condenser and the evaporator. While an array of varying patterns both on the evaporator and the condenser side of the VC have been studied, a condenser design comprising of wettability patterns as developed by Ghosh et al. [6] complemented with low Laplace pressure condensate wells was found to be the most effective. The evaporator on the other hand, had a wettability pattern designed for continuous pumpless transport (via implementation of superhydrophilic wedges on a hydrophobic background) of the working fluid toward the heated area. However, experimentation of such wick-free vapor chamber is often hindered by the reliability of the experimentation apparatus that maintains system vacuum, or by geometrical constraints for the developed chamber.

Herein, we develop an accurate system for extensive experimentation of wick-free VC. We shift from our earlier approaches of having an inlet/outlet port via a gasket, and use permanent ports designed on the top plate of the VC. Such an approach eliminates any leakage of NCGs into the systems, while allowing the chamber to remain in relative vacuum for extended periods of experimentation. The new design of the experimental setup also allows for the VCs to be thinner, without any restrictions on the gasket dimensions, as it only functions to maintain vacuum and the vapor space in between the evaporator and the condenser. We delve into the development of thin wick-free VCs with our new setup, and reduce the overall thickness of the VC to 3 mm. Furthermore, we incorporate wettability patterns corresponding to the best performing evaporator and condenser as reported by Damoulakis and Megaridis [11]. Thereafter, we investigate the effect of patterns in such thin wick-free VCs, by comparison against VCs with uniform surfaces and no wettability patterns.

## II. Materials & Methodology

### A. Vapor Chamber Fabrication

A wick-free vapor chamber with laser etched patterns and functionalized surfaces is fabricated for spreading heat from a ceramic heater (area 0.9072 cm$^2$). The VC consists of three components, namely, an evaporator plate, a condenser plate, and a gasket, which maintains an air-tight seal, while also acting as a separator in between the two plates and maintaining the vapor space. This approach is not new, and provides modularity for experimentation, allowing researchers to swiftly change components and make modifications for development of better devices [8, 9]. The VC has a total thickness of 3 mm, which comprises of 1 mm-thick evaporator and condenser plates, while having total vapor space of 5 cm × 5 cm × 1 mm. For benchmarking purposes, a VC comprising of a uniformly superhydrophobic condenser and a uniformly superhydrophilic evaporator (as shown in Figure 1 – A) was developed and its thermal performance also evaluated. For further characterization of the device's performance, a patterned condenser along with a patterned evaporator were used (as shown in Figure 1 – B).

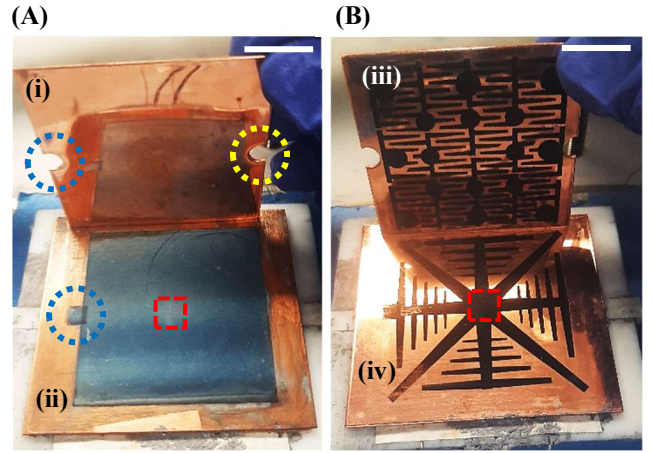

Fig. 1. Two different wick-free VCs were used to evaluate thermal performance of the developed system: **(A)** A non-patterned VC with **(i)** uniformly hydrophobic condenser, and **(ii)** uniformly superhydrophilic evaporator. **(B)** A wettability-patterned VC comprising of **(iii)** a patterned condenser, and **(iv)** a patterned evaporator. Two ports on the condenser side of the VC allow for degassing the system (marked with a yellow dashed circle in (A)) and for charging the system with the working fluid. Red-dashed squares near the centre of each evaporator plate, approximately mark the location of the heater underneath both plates. The white scale bars on the top right of both frames represent 2 cm of length.

Both the evaporator and the condenser comprise of 1 mm-thick copper plates (Cu-101, thermal conductivity of 390.8 W/mK, McMaster-Carr). Each plate, as received, was diced into smaller pieces of 6.5 cm × 7 cm for the evaporator and 5.3 cm × 5.3 cm for the condenser. Two holes for the charging inlet (see blue dashed circle in Figure 1 – A) and the degassing unit (see yellow dashed circle in Figure 1 – A) were punched on the condenser side of the VC. The cut-out plates were then subjected to ultrasonication first in ethanol (purity > 98%) and then in distilled water, both for 15 minutes. Thereafter, the plates were dried by blowing nitrogen over their surface. A two-step process was implemented for functionalization of the surfaces. First, the surfaces were coated with Teflon and then subjected to selective etching of the Teflon coating to expose hydrophilic pathways on a hydrophobic background. For coating the surface with Teflon, the dried cleansed surfaces were spin-coated at 2000 rpm for 20 seconds with Teflon AF (AF-2400, Amorphous Fluoroplastics Solution, Chemours Co.). The surface was then cured in a single-zone tube furnace (Lindberg, Blue-M-HTF55322c) in a reducing environment consisting of 10% Hydrogen and 90% Argon, in three stages viz., 80 °C, 180 °C and 260 °C, each for 10 minutes, to completely cure the surface. The resulting surface was uniformly hydrophobic, and reports a static contact angle of 118° ± 2° (as shown in Figure 2 – B), when measured with an optical goniometer.

*Patterning and laser etching*: The evaporator and the condenser were designed to be either of uniform wettability, or to have designated patterns, that help in condensation drainage and continuous supply of charging fluid (water) to the heated area. Such mechanism of pumpless water transport has been analyzed in earlier research by Ghosh et al. [6] to improve condensation and by Damoulakis and Megaridis [11] for development of wick-free VCs. A laser marking system (TYKMA Electrox EMS – 400) was used at 40 % power output, 20kHz intensity and with a traverse speed of 200 mm/s, to either etch uniformly or etch designated patterns onto the evaporator surface. Such etching imparts surface roughness, while also removing any coatings to expose the bare copper surface. For development of the VC with no patterns, a uniformly etched 5 cm × 5 cm square area was imparted on the evaporator plate. An additional small square area corresponding to the blue circle in Figure 1 – A, was etched directly below the charging port (also marked in blue), to ease the entry of water from the port into the chamber.

After the surface had been etched with the laser, it was subjected to a wet chemical process, to turn the etched areas superhydrophilic. The Teflon-coated areas were chemically inert and did not show any change of surface properties, even after exposure to chemicals. The plate sample was immersed into a solution of 2.5 molar NaOH (Sodium hydroxide, Sigma Aldrich – 415413 – 500mL) and 0.1 molar $(NH_4)_2S_2O_8$ (Ammonium persulfate ≥ 98%, Sigma Aldrich – MKCF3704). This chemical treatment was performed at room temperature and formed copper hydroxide nano-hairs (as seen from the scanning electron micrograph in Figure 2 – C) on the laser-etched surface, resulting in a superhydrophilic surface [12]. Thus, the generated surface has superhydrophilic tracks on a Teflon-coated hydrophobic background.

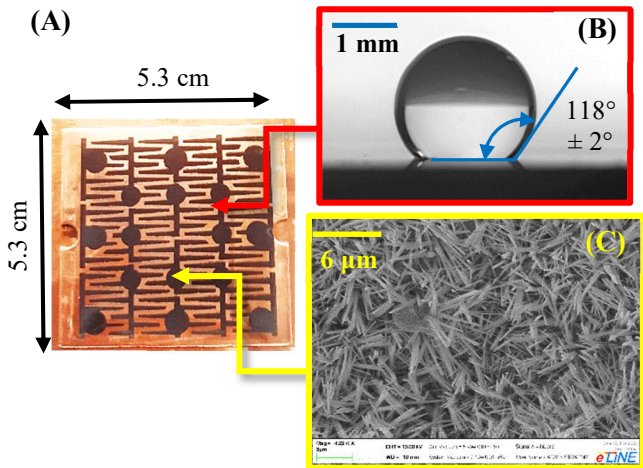

Fig. 2. (A) A patterned condenser used for experimentation in the wick-free vapor chamber. The patterns comprise of super-hydrophilic tracks (black) on a hydrophobic (copper colored) background. (B) The static contact angle of the hydrophobic (Teflon-coated copper) surface can be estimated using optical goniometry, as shown in the top right (red inset box). The laser-etched hydrophilic pathways are subjected to chemical treatment, which grows (C) copper hydroxide nano-hairs (as seen under a scanning electron microscope) and renders these pathways superhydrophilic.

### B. Choice of pattern for wick-free VC

It is important to note here that the thermal performance of the VC is strongly influenced by the choice of pattern on both the condenser and the evaporator of the system. While hydrophobic surfaces have been known to improve condensation by almost an order of magnitude in pure steam conditions, bio-inspired wettability patterns add further to the efficiency and enhancement of heat transfer rates [6]. The choice of a condenser pattern needs to facilitate complementary evaporator designs, that would drain the water from specific points on the condenser (low-pressure wells that collect the condensate), and move them towards the heated area for evaporation (two-phase) heat transfer. Herein, the choice of patterns to carry out the VC experiments is not arbitrary, but corresponds to the complementary designs resulting in the best thermal resistance, as previously reported by Damoulakis and Megaridis [11]. For a better understanding of how the condenser drains into the evaporator, and why such patterns result in decreased VC thermal resistance, the reader is referred to the aforementioned works of Damoulakis and Megaridis, where multiple patterns and their thermal performance were compared in wick-free VCs.

### C. Experimental Setup

The experimental setup shown in Figure 3 was designed for thermal performance evaluation of the present VCs. The vapor chamber was assembled by arranging the copper plates and the gaskets, such that they were aligned on top of the heater and with aligned charging and vacuum ports. Earlier VC experimentation employed inlet/outlet ports through the gaskets, thus imposing minimum size restrictions on the total thickness of the vapor chamber [8, 9, 11]. In the present design, the inlet/outlet ports are through the cold plate (see Figure 3 – C & D), thus removing any such restrictions, and allowing experimentation of thin VCs.

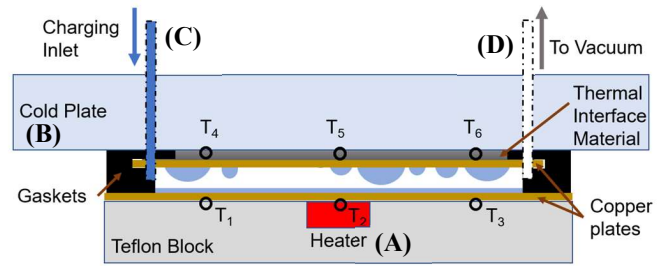

Fig. 3. Schematic depiction of experimental setup developed for testing the present wick-free VCs, which comprise of surface-functionalized copper plates, spaced by gaskets, which also maintained the system vacuum. The VC sits atop a heater (A), and is held together by a cold plate (B) which is secured onto the bench top, over which the entire setup is housed. Strategically located charging inlet (C) and vacuum ports (D), go through the cold plate, and allow for testing of thin chambers without severe size constraints.

The vapor chamber was placed on top of a ceramic resistive heater (Component General, CPR-375-1) of dimensions 9.525 mm × 9.525 mm × 1.016 mm, which generated heat as the applied voltage across the heater was varied from 10 – 60 V (supplied from a voltage regulation system: Volteq-HY10010EX). The heater was in turn housed in a Teflon block (8735 K67 McMaster-Carr), so that the generated heat can only be rejected via the VC. A cold plate attached to the top of the VC could be set to any temperature of the user's choice, to remove heat from the VC in a controlled fashion. The cold plate also held the VC in place, secured onto the work bench, and maintained pressure so that the gasket (Viton Fluoroelastomer, with a durometer hardness of 75A, McMaster-Carr) maintained proper contact between the evaporator and the condenser. Strategically placed thermocouples (as shown in Figure 3: $T_1$ – $T_6$) (T-Type, Omega, bead diameter – 0.13mm) both on the evaporator and condenser side of the VC, relayed information to a data acquisition platform (Omega DAQ, USB 2400), which in turn recorded data with a frequency of 1Hz (sampling rate) to a connected PC. Thermal gap pads (Bergquist Company, thickness of 0.25 mm, thermal conductivity 3 W/mK) were used to maintain proper thermal contact without air gaps between the VC, the heater, the cold plate and the thermocouples.

*D. Experimental Procedure*

A standard experimental procedure was followed for performing all experimental trials. After the setup had been assembled (as shown in Figure – 3), the VC was degassed by switching on the vacuum pump (Alcatel Annecy – 2008A), until a pressure of ~6.5kPa was reached, and almost all NCGs from the vapor chamber had been removed. The vapor chamber was left for more than an hour in its evacuated state, to ensure that the setup could hold the vacuum without any leaks, for the duration of the experiment. Thereafter, the VC was charged with a pre-determined amount of the working fluid, and all ports connecting the vapor chamber were closed. Each experimental trial at a particular heat input was carried out for 5 minutes, and the temperature data from the last 60 seconds was used to calculate the relevant parameters. For each set of experiments, the power was continuously increased until the temperature over the heater reaches 120 °C, beyond which the heater resistance values begin to fluctuate, or until the system reached a thermal dry-out (which was indicated by a sharp increase of the evaporator thermocouple reading $T_2$). For all experimental trials, the cold plate was maintained at a constant temperature of 19 °C.

*E. Performance Metrics and Data Reduction*

The data generated from each VC was stored in real-time on a PC via the DAQ, and later used for evaluating the VC performance metrics. For evaluating VC performance, the overall device thermal resistance was calculated from the temperature data. The thermal resistance $R_{total}$ is the ratio of the difference between the heater and condenser temperatures over the power input to the system, i.e.

$$R_{total} = \frac{\Delta T}{Q} = \frac{T_{heater} - T_{cond}^{avg}}{Q_{in}} \quad (1)$$

where $Q_{in}$ is the heat input to the system, and $T_{heater}$ and $T_{cond}^{avg}$ are the temperatures over the heater, and the average of the thermocouple values on the condenser side of the system, respectively. It is also important to note another parameter which strongly influences the performance of the VC, namely its charging ratio (*CR*). The charging ratio of the VC is defined as the ratio of the volume of the working fluid to the volume of the vapor space, i.e.,

$$CR = \frac{Volume\ of\ working\ fluid}{Vapor\ Space} \quad (2)$$

A lower *CR* can result in thermal dry-out, whereas a higher *CR* can result in increased thermal resistance of the system. For all parameters estimated during the course of this study, a Gaussian error propagation – as carried out in our earlier works- have been implemented to estimate experimental errors [8, 9].

III. RESULTS & DISCUSSION

Experiments were first carried out on a VC comprising of a uniformly hydrophobic condenser (mirror-finished copper coated with Teflon) and a uniformly superhydrophilic evaporator. Tests were run for three different charging ratios (*CR*s) and the thermal resistance for each CR was calculated for varying power inputs, as shown Figure 4. Lower $R_{total}$ values were observed for low *CR*s and at lower power inputs. At higher power inputs, this trend reverses and higher *CR*s result in lower thermal resistances. This trend is logical and is explained by the fact that the VCs would move towards thermal dry-outs at higher power input; only a greater quantity of the working fluid can keep the system running efficiently. On the other hand, for lower power inputs, a higher *CR* results in a thicker layer of working fluid on the evaporator side, which adds to the thermal resistance, thereby decreasing overall device performance. For this particular configuration of the VC, the lowest thermal resistance (~0.4 K/W) values were observed at a power input of 22.5 W, and a *CR* of 24 %.

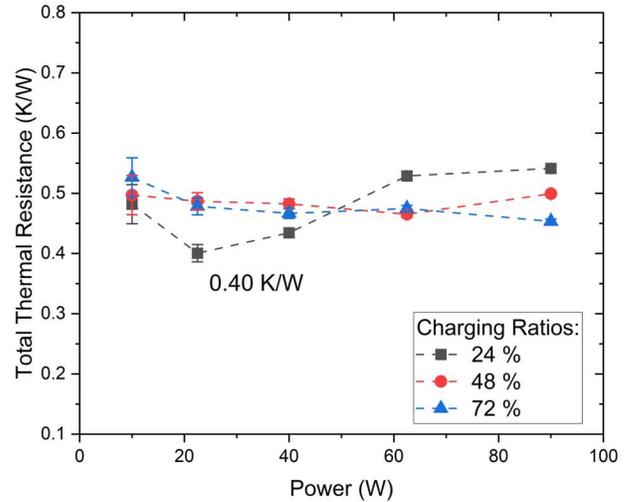

Fig. 4. Total thermal resistance vs. power input to the wick-free VC (comprising of a uniformly hydrophobic condenser and a uniformly superhydrophilic evaporator), at varying water charging ratios. The lowest thermal resistance of 0.40 K/W is observed at a power input of 22.5 W and *CR* of 24%.

Similar experiments were also run on a VC comprised of a patterned condenser and a patterned evaporator. The patterns of the evaporator and the condenser are complementary, and designed such that the wells of the condenser drain into specific locations on the evaporator. The design and fabrication methodology implemented for the design of such a VC has been explained earlier. Tests were run for three different charging ratios and the thermal resistances of each VC were calculated for varying power inputs, as plotted in Figure 5.

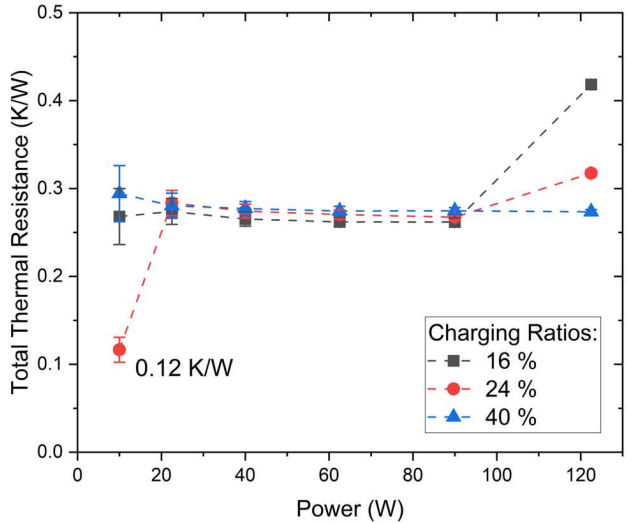

Fig. 5. Total thermal resistance vs. power input to the wick-free VC (comprising of a wettability-patterned condenser and a patterned evaporator) for varying fluid charging ratios. The lowest thermal resistance of 0.12 K/W is observed at a power input of 10 W, and at a CR of 24%.

For the patterned VC, significant differences were observed from the benchmark case of the non-patterned vapor chamber. The lowest thermal resistance (~0.12 K/W) was observed at a power input of 10 W, and a *CR* of 24 %. The thermal resistances of the VC for all *CR*s and at different power inputs decrease after patterning of the evaporator and the condenser. The patterned VC performs similar to the non-patterned VC at higher power inputs; higher *CR*s perform better, while lower *CR*s show better performance at lower power inputs. Here, it is interesting to note that the patterning of the evaporator renders the *VC* thermal resistance independent of the variation of *CR*. While this outcome is non-intuitive, it can be explained by the fact that the patterns on the evaporator side of the VC confine the liquid within the superhydrophilic tracks, unlike uniformly superhydrophilic evaporators, which spread the liquid all over the surface. Such confinement ensures that an adequate quantity of liquid is always present on the hot spot, thus providing effective cooling without dry-out. At the same time, the condenser drains onto the evaporator patterns, thus ensuring the working fluid cycles through evaporation and condensation, while keeping the vapor space unblocked, and allowing unhindered flow of vapor from the evaporator to the condenser.

## IV. CONCLUSIONS

We report a thin wick-free vapor chamber with an overall thickness of 3 mm, a 40% reduction from our earlier wick-free VC design. A specialized wick-free VC test platform is developed and used for testing of thin VCs. The developed chamber used charging and degassing ports on the condenser side of the VC, ensuring a proper seal during experimentation. The setup was also equipped with a cold plate that could maintain the condenser at any desired temperature. This setup was used to investigate the thermal performance of the wick-free VC, while comparing performance metrics for patterned and non-patterned vapor chamber designs.

Two distinctive set of experiments were carried out. In the first scenario, thermal performance evaluation of a VC with a uniformly hydrophobic condenser and a uniformly superhydrophilic evaporator (no patterns on either side) were performed. The lowest thermal resistance of 0.40 K/W was observed at a *CR* of 24% and for a power input of 22.5 W. An even lower thermal resistance of 0.12 K/W (*CR*: 24%, power input: 10 W) was observed for a VC with patterned evaporator and patterned condenser. In the case of the VC with no patterns, a reversal in the relation between the thermal performance and the *CR* is seen at higher powers. The thermal resistances are seen to be strongly influenced by the wettability patterns, which almost render the VCs independent of the choice of *CR*. This is attributed to the confining and fast transport properties of wettability patterning, which has been previously known to enhance heat transfer in condensation [6].

Overall, the results show that the performance of the present thin VC is comparable with that obtained in earlier research by Damoulakis and Megaridis [11]. We also conclusively demonstrate that wettability patterning in wick-free vapor chambers improves performance, while allowing VCs to work at even higher power inputs without reaching dry-outs. Further research to investigate the development of thin wick-free VCs or understand how the pattern shapes and their relationship with *CR* influence thermal performance of the VCs remains attractive for future directions.

ACKNOWLEDGMENT

This material is based upon research supported by, or in part by , the US Office of Naval Research under award number N14-20-1-2025 to the University of Nebraska, Lincoln and a subaward to the University of Illinois Chicago. The authors thank J. Rodriguez (Assistant Director) and T. Bruzan (Laboratory Mechanic) at the UIC College of Engineering Scientific Instrument/Machine shop for machining the samples. The authors also thank the UIC College of Engineering, Nanotechnology Core Facility (NCF), for further development and characterization of the samples used in this work.